\documentstyle[12pt]{article}
\title{Radiative tail in $\pi_{e2}$ decay and some comments on
$\mu$--$e$ universality}

\author{E.A.~Kuraev}

\date{}

\newcommand{\eps}{\varepsilon}
\newcommand{\dd}{\mbox{d}}
\begin{document}

\maketitle

\begin{center}
{\em Bogoliubov Laboratory for Theoretical Physics, JINR,
Dubna, 141980, Russia\/}
\end{center}

\begin{abstract} \noindent
The result of lowest-order perturbation theory calculations of the
photon and positron spectra in radiative $\pi_{e2}$ decay are generalized
to all orders of perturbation theory using the structure-function
method. An additional source of radiative corrections to the ratio of
the positron and muon channels of pion decay, due to emission of
virtual and real photons and pairs, is considered. It depends on details
of the detection of the final particles and is large enough to be
taken into account in theoretical estimates with a level of accuracy
of $0.1\%$.
\end{abstract}

\vspace{.5cm}


\thispagestyle{empty}
%\vspace{.2cm}
\newpage
\setcounter{page}{1}

As a first step in the calculation of the spectra of radiative pion decays
we reproduce the results obtained by Berman and Kinoshita~[1],
treating the pion as a point-like particle.
Kinoshita~[1] calculated the positron energy spectrum
in radiative pion decay:
\begin{eqnarray}
&& \frac{\dd\Gamma}{\Gamma_0\dd y}=\frac{\alpha}{\pi}\biggl[
\frac{1+y^2}{1-y}(L-1) - 1 + y - \frac{1}{2}(1-y)\ln(1-y)
 + \frac{1+y^2}{1-y}\ln y\biggr],\\ \nonumber
&& \quad y_{\rm min}\leq y\leq 1+\frac{m_e^2}{m_{\pi}^2},
\end{eqnarray}
where $y=\frac{2\eps_e}{m_{\pi}}$ is the positron energy fraction, $\eps_e$
is the positron energy (here and below we have in mind the rest frame of
the pion),
$L=\ln(m_{\pi} m_e)=5.6$ is the ``large logarithm'', and
$m_{\pi}$, $m_e$ are the masses of the pion and positron.
The quantity
\begin{eqnarray}
\Gamma_0=\frac{G^2|V_{ud}|^2}{8\pi}f_{\pi}^2m_e^2m_{\pi}
\biggl(1-\frac{m_e^2}{m_{\pi}^2}\biggr)^2=2.53\cdot 10^{-14}\ \mbox{MeV},
\end{eqnarray}
is the total width of $\pi_{e2}$ decay, calculated in the Born
approximation.

We will now calculate the photon spectrum. Consider first the emission
of a soft real photon. The corresponding contribution to the total width
may be obtained by the standard integration of the differential widths:
\begin{eqnarray}
\frac{\dd\Gamma^{\rm soft}}{\Gamma_0}=-\frac{\alpha}{4\pi^2}
\int\frac{\dd^3 k}{\omega}\;\left(\frac{P}{Pk}
- \frac{p_e}{p_ek}\right)^2
\Bigg|_{\omega\leq\Delta\eps\ll m_{\pi}/2},
\end{eqnarray}
where $P,p_e,k$ are the four-momenta of the pion, positron, and photon,
respectively, $P^2=m_{\pi}^2$, $p_e^2=m_e^2$, $k^2=\lambda^2$,
and $\lambda$ is the photon mass. The result has the form
\begin{eqnarray}
\frac{\Gamma^{\rm soft}}{\Gamma_0}&=&\frac{\alpha}{\pi}
\biggl[-b(\sigma)\ln\frac{2\Delta\eps}{\lambda}
+ 1 - \frac{1+\sigma}{2(1-\sigma)}\ln\sigma \\ \nonumber
&-& \frac{1+\sigma}{4(1-\sigma)}\ln^2\sigma -
\frac{1+\sigma}{1-\sigma}\mbox{Li}_2(1-\sigma)\biggr],
\end{eqnarray}
where
\begin{eqnarray}
b(\sigma)=\frac{1+\sigma}{1-\sigma}\ln\sigma+2, \quad
\sigma=\frac{m_e^2}{m_{\pi}^2},\quad
\mbox{Li}_2(x)=-\int\limits_{0}^{x}\frac{\dd t}{t}\ln(1-t).
\end{eqnarray}

Consider now the hard photon emission process
\begin{eqnarray}
\pi^+(P) \to e^+(p_e)+\nu_e(p_{\nu})+\gamma(k).
\end{eqnarray}
The standard procedure of final-states summing of the squared modulus
of its matrix element and integration over the neutrino phase volume
leads to the spectral
distribution over the photon energy fraction $x=2k^0/m_{\pi}$:
\begin{eqnarray}
\frac{\dd\Gamma}{\Gamma_0\dd x}&=&\frac{\alpha}{2\pi}
\frac{x(1-x-\sigma)}{(1-\sigma)^2}
\biggl[-\frac{4(1-\sigma)}{x^2}-\frac{1}{1-x} \\ \nonumber
&+& \frac{1}{x(1-x-\sigma)} \biggl(\frac{1}{x}(1+(1-x)^2)
+ 2\sigma-2\frac{\sigma^2}{x}\biggr)\ln\frac{1-x}{\sigma}\biggr].
\end{eqnarray}
Further integration of this spectrum give the result
\begin{eqnarray}
\int\limits_{x_{\rm min}}^{1-\sigma}\frac{\dd\Gamma}{\Gamma_0\dd x}\dd x
&=& \frac{\alpha}{2\pi}\biggl[-2b(\sigma)\ln\frac{1-\sigma}{x_{\rm min}}
- 2\frac{1+\sigma}{1-\sigma}\mbox{Li}_2(1-\sigma) \\ \nonumber
&+& \frac{3(1-2\sigma)}{2(1-\sigma)^2}\ln\sigma
+ \frac{19-25\sigma}{4(1-\sigma)}\biggr],
\quad x_{\rm min}=\frac{2k^0_{\rm min}}{m_{\pi}}.
\end{eqnarray}
Putting $k^0_{\rm min}=\Delta\eps$ in this formula and adding the soft
photon contribution, we obtain (in agreement with Kinoshita's
1959 result) the contribution to the width from the inner
bremsstrahlung of a point-like pion:
\begin{eqnarray}
\frac{\Gamma_{IB}}{\Gamma_0}&=&\frac{\alpha}{\pi} \biggl\{b(\sigma)
\biggl[\ln\frac{\lambda}{m_{\pi}} - \ln(1-\sigma)-\frac{1}{4}\ln\sigma
+ \frac{3}{4}\biggr] \\ \nonumber
&-& 2\frac{1+\sigma}{1-\sigma}\mbox{Li}_2(1-\sigma)
- \frac{\sigma(10-7\sigma)}{4(1-\sigma)^2}\ln\sigma
+ \frac{15-21\sigma}{8(1-\sigma)}\biggr\}.
\end{eqnarray}

Now return to the positron spectrum. The contributions to it
containing the large logarithm $L$ may be associated with the known
kernel of the Altarelli--Parisi--Lipatov evolution equation (see[2]):
\begin{eqnarray}
P^{(1)}(y)=\lim\limits_{\Delta\to 0}\biggl[
\frac{1+y^2}{1-y}\theta(1-y-\Delta)
+(\frac{3}{2}+2\ln\Delta)\delta(1-y)\biggr]
=\biggl(\frac{1+y^2}{1-y}\biggr)_+.
\end{eqnarray}
Using the factorization theorem, we may generalize this spectrum
to include the leading logarithmic terms in all orders of perturbation
theory. This may be done in terms of structure functions $D(y,\sigma)$~[2].
In the case of the photon spectrum the function $D(1-x,\sigma)$ appears.
The function
$D(y,\sigma)$ describes the probability of finding a positron with energy
fraction $y$ inside the initial positron. It may be present in the form of
a sum
 of non-singlet and singlet contributions,
$D=D^{\gamma}+D^{e^+e^-}$.
Iteration of the evolution equations gives
\begin{eqnarray}
D(y,\sigma)=\delta(1-y)+P^{(1)}(y)\gamma
+\frac{1}{2}(P^{(2)}(y)+P^{e^+e^-}(y))\gamma^2+\dots,
\end{eqnarray}
where
\begin{eqnarray}
\gamma&=&-3\ln(1-\frac{\alpha}{3\pi}(L-1)), \nonumber \\
P^{(2)}(y)&=&\int\limits_{y}^{1}\frac{\dd
t}{t}P^{(1)}(t)P^{(1)}(\frac{y}{t})
=\lim\limits_{\Delta \to 0}\biggl\{\biggl[(2\ln \Delta+\frac{3}{2})^2
\nonumber \\
&-& \frac{2\pi^2}{3}\biggr]\delta(1-y)+2\biggl[\frac{1+y^2}{1-y}(2\ln(1-y)
- \ln y+\frac{3}{2}) \nonumber  \\
&+& \frac{1}{2}(1+y)\ln y-1+y\biggr]\theta(1-y-\Delta)\biggr\}, \\
\nonumber
P^{e^+e^-}(y)&=&\frac{2}{3}P^{(1)}(y)+\frac{(1-y)}{3y}(4+7y+4y^2)+2(1+y)\ln
 y.
\end{eqnarray}
It is convenient to use the smoothed form of them:
\begin{eqnarray}
D^{\gamma}(y) &=& \frac{1}{2}b(1-y)^{\frac{1}{2}b-1}
\biggl[1+\frac{3}{2}b-\frac{1}{48}b^2(\frac{2}{3}L+\pi^2-\frac{47}{8})\biggr
]
\\ \nonumber
&-& \frac{1}{4}b(1+y)+\frac{1}{32}b^2[4(1+y)\ln\frac{1}{1-y}
+ \frac{1+3y^2}{1-y}\ln\frac{1}{y}-5-y] + {\cal O}(b^3),\\ \nonumber
D^{e^+e^-}(y) &=& \frac{1}{3}\biggl(\frac{\alpha}{\pi}(L-\ln(1-y)
- \frac{5}{6})\biggr)^2(1-y)^{\frac{1}{2}b-1}\;
(1+y^2+\frac{1}{3}b(L-\ln(1-y) \\ \nonumber
&-& \frac{5}{6}))
+\frac{1}{96}b^2\biggl[\frac{1-y}{y}(4+7y+4y^2)+6(1+y)\ln y\biggr]
+{\cal O}(b^3),\\ \nonumber
b &=& \frac{4\alpha}{\pi}(L-1).
\end{eqnarray}
The expressions for spectra are as follows:
\begin{eqnarray}
\frac{\dd\Gamma}{\Gamma_0\dd y}&=&D(y,\sigma)[1+\frac{\alpha}{\pi}K_e(y)],
\\ \nonumber
K_e(y)&=&-1+y-\frac{1-y}{2}\ln(1-y)+\frac{1+y^2}{1-y}\ln y,\quad
y=\frac{2\eps_e}{m_{\pi}}, \\ \nonumber
\frac{\dd\Gamma}{\Gamma_0\dd x}&=&D(1-x,\sigma)[1+\frac{\alpha}{\pi}
K_{\gamma}(x)], \\ \nonumber
K_{\gamma}(x)&=&x+\frac{1+(1-x)^2}{x}\ln(1-x),\quad
x=\frac{2\omega}{m_{\pi}}.
\end{eqnarray}

Let us discuss the contribution of the inelastic processes considered above
to the ratio of the widths of the positron and muon modes of pion
decay, $R_{\pi_{l2}}$:
\begin{eqnarray}
R_{\pi l2}=\frac{\Gamma(\pi\to e\nu)+\Gamma(\pi\to e\nu\gamma)}
{\Gamma(\pi\to \mu\nu)+\Gamma(\pi\to \mu\nu\gamma)}\, .
\end{eqnarray}
Close attention was paid to this quantity some years ago [3,4], but the
corrections for emission processes in higher orders of perturbation
theory were not taken into account.
Keeping in mind that the quantity $P^{(1)}(y)$ has the property
\begin{eqnarray}
\int\limits_0^1\dd y\; P^{(1)}(y)=0,
\end{eqnarray}
we make the important observation that as long as an experiment is
proceeding in such a way that no cuts are imposed on the positron energy,
then no large logarithmic contributions appear. However, if the cuts
are such that the $y$ integration is restricted or convoluted with a
$y$-dependent function, some terms proportional to the large logarithm
$L$ will remain. We now suggest that there exists some minimum energy
$\eps_{\rm th}$ for detection of the positron. An additional contribution
(not considered in [4])
appears:
\begin{eqnarray}
\frac{\Delta R}{R_0}&=& -\frac{\alpha}{\pi}\int\limits_0^{x_{\rm th}}\dd
x\;
\frac{1+x^2}{1-x}(L-1) \\ \nonumber
&+& \frac{\alpha}{\pi}\int\limits_{x_{\rm th}}^1\dd x\; D(x,\sigma)K_e(x),
\qquad
x_{\rm th}=\frac{2\eps_{\rm th}}{m_{\pi}},
\end{eqnarray}
where
\begin{eqnarray}
R_0=\frac{m_e^2}{m_{\mu}^2}\,\frac{(1-m_e^2/m_{\pi}^2)^2}
{(1-m_{\mu}^2/m_{\pi}^2)^2} = 1.28347\cdot 10^{-4}.
\end{eqnarray}
For typical values $x_{\rm th}=0.1$ this additional contribution will have
a magnitude of order $10^{-3}$ and should be taken into account in
calculations for accuracy at the $0.1\%$ level.

\subsection*{Acknowledgement}
The author is grateful to the RFFI foundation for grant
${\cal N}^{\circ}$~96-02-17512. He is also thankful to V.~Gordeev
and A.~Arbuzov for discussions and help.


\begin{thebibliography}{99}

\bibitem{r1}
S.M.~Berman, Phys. Rev. Lett. {\bf 1}, 468 (1958); \\
T.~Kinoshita, Phys. Rev. Lett. {\bf 2}, 477 (1959).

\bibitem{r2}
E.A.~Kuraev and V.S.~Fadin, Sov. J. Nucl. Phys. {\bf 41}, 466 (1985).

\bibitem{r3}
D.~Bryman, Comments Nucl. Part. Phys. {\bf 21}, 101 (1993).

\bibitem{r4}
W.~Marciano and A.~Sirlin, Phys. Rev. Lett. {\bf 71}, 3629 (1993).

\end{thebibliography}
\end{document}